\documentclass[10pt,a4paper]{article} 
\usepackage{jcappub}
\pdfoutput=0

\newcommand{\bdv}[1]{{\bf #1}}
\newcommand{\bdi}[1]{\hbox{\boldmath{$#1$}}}
\newcommand{\oo}{v}         
\newcommand{\cc}{\lambda}   
\newcommand{\zz}{z}      
\newcommand{\zh}{\bar\zz}      
\newcommand{\ttt}{\theta}   
\newcommand{\pp}{\phi}     
\newcommand{\rhat}{\hat n}   
\newcommand{\nhat}{\bdi{\rhat}}  
\newcommand{\NC}{\mathbb{C}}   
\newcommand{\CK}{\hat k}         
\newcommand{\NN}{N}              
\newcommand{\CG}{\hat g}         
\newcommand{\CU}{\hat u}         
\newcommand{\phat}{\hat\pp}  
\newcommand{\that}{\hat\ttt}  
\newcommand{\thatv}{\bdi{\that}}  
\newcommand{\phatv}{\bdi{\phat}}  
\newcommand{\rbar}{\bar r}
\newcommand{\UU}{u}
\newcommand{\VV}{\mathcal{U}}  
\newcommand{\SVV}{U}           
\newcommand{\VVV}{U}           
\newcommand{\VGI}{V}           
\newcommand{\gv}{v}            
\newcommand{\sv}{v}            
\newcommand{\GG}{\delta\Gamma}    
\newcommand{\Dcc}{\Delta\cc}
\newcommand{\DT}{\Delta\tau}    
\newcommand{\dTau}{\Delta\tau}
\newcommand{\drr}{\delta r}    
\newcommand{\dtt}{\delta\ttt}  
\newcommand{\dpp}{\delta\pp}   
\newcommand{\dg}{\delta g}
\newcommand{\dV}{\delta V}
\newcommand{\dm}{\delta_m}
\newcommand{\dtp}{\delta_m^{t_p}}
\newcommand{\dobs}{\delta_g^\up{obs}}
\newcommand{\dobsTHR}{\delta_{g,\up{3D}}^\up{obs}}
\newcommand{\dobsTWO}{\delta_{g,\up{2D}}^\up{obs}}
\newcommand{\nobsTWO}{n_{g,\up{2D}}^\up{obs}}
\newcommand{\dv}{\delta_v}
\newcommand{\av}{\alpha_v}
\newcommand{\pN}{\phi_N}  
\newcommand{\vN}{v_N}     
\newcommand{\TT}{T}            
\newcommand{\PZ}{\mathcal{P}_z}  
\newcommand{\PR}{\Upsilon}     
\newcommand{\cA}{\mathcal{P}}
\newcommand{\cB}{\mathcal{R}}
\newcommand{\SP}{\mathcal{S}}
\newcommand{\RW}{N}             
\newcommand{\WW}{\Xi}           
\newcommand{\RR}{\mathcal{R}}   
\newcommand{\WM}{\hat W}        
\newcommand{\TC}{\mathcal{T}}   
\newcommand{\ST}{\mathcal{M}}   
\newcommand{\up}[1]{{\rm #1}}
\newcommand{\beeq}{\begin{equation}}
\newcommand{\eneq}{\end{equation}}
\newcommand{\bear}{\begin{eqnarray}}
\newcommand{\enar}{\end{eqnarray}}
\newcommand{\la}{\langle}
\newcommand{\ra}{\rangle}
\newcommand{\OM}{\Omega_m}
\newcommand{\AVE}[1]{\langle#1\rangle}
\newcommand{\kvec}{\bdv{k}}
\newcommand{\xvec}{\bdv{x}}
\newcommand{\Kang}{\bdv{\hat k}}
\newcommand{\HH}{\mathcal{H}}   
\newcommand{\gbar}{\bar g}      
\newcommand{\dz}{\delta z}           
\newcommand{\dL}{\mathcal{D}_L}      
\newcommand{\ddL}{\delta\mathcal{D}_L} 
\newcommand{\dA}{\mathcal{D}_A}      
\newcommand{\BB}{\mathcal{B}}    
\newcommand{\CC}{\mathcal{C}}  
\newcommand{\dT}{\delta\tau}        
\newcommand{\ax}{\alpha_{\chi}}     
\newcommand{\px}{\varphi_{\chi}}    
\newcommand{\cp}{\varphi_v}         
\newcommand{\vx}{v_{\chi}}          
\newcommand{\pv}{\Psi}              
\newcommand{\dnu}{\delta\nu}        
\newcommand{\dea}{{\delta n}}       
\newcommand{\dnug}{\delta\nu_\chi}  
\newcommand{\deag}{{\delta n_\chi^\alpha}}      
\newcommand{\CCG}{\mathcal{G}}      
\newcommand{\dzg}{\dz_\chi}         
\newcommand{\kag}{\mathcal{K}}      
\renewcommand{\AA}{\mathcal{A}}

\title{Relativistic Effect in Galaxy Clustering}

\author[a,b]{Jaiyul Yoo}

\affiliation[a]{Center for Theoretical Astrophysics and Cosmology,
Institute for Computational Science, University of Z\"urich, Switzerland}
\affiliation[b]{Physics Institute, University of Z\"urich, 
Winterthurerstrasse 190, CH-8057, Z\"urich, Switzerland}
\affiliation{\bigskip(submitted 18 June 2014; accepted 10 September 2014)\bigskip}

\emailAdd{jyoo@physik.uzh.ch}

\abstract{The general relativistic description of galaxy clustering provides
a complete and unified treatment of all the effects in galaxy clustering
such as the redshift-space distortion, gravitational lensing, Sachs-Wolfe
effects, and their relativistic effects. In particular, the relativistic
description resolves the gauge issues in the standard Newtonian description
of galaxy clustering by providing the gauge-invariant expression 
for the observed galaxy number density. The relativistic effect in galaxy
clustering is significant on large scales, in which dark energy models or
alternative theories of modified gravity deviate from general relativity.
In this paper, we review the relativistic effect in galaxy clustering
by providing a pedagogical derivation of the relativistic formula and
by computing the observed galaxy two-point statistics. The relativistic 
description of galaxy clustering is an essential tool for testing general
relativity and probing the early Universe on large scales in the era of
precision cosmology.}

\renewcommand{\beforetochook}{\clearpage}  
\toccontinuoustrue    

\begin{document}
\maketitle
\flushbottom

\section{Introduction}
\label{sec:intro}

To understand the nature of dark energy and to probe the early Universe,
a large number of galaxy surveys are operational and even a larger number
of surveys are planned for a near future. These current and future galaxy
surveys will deliver high precision measurements of galaxy clustering,
providing enormous statistical power to solve the issues in the standard
model of cosmology. However, high precision measurements in the
upcoming galaxy surveys simultaneously
bring in new challenges, setting the level of accuracy that theoretical
predictions are obliged to meet. Given these strict requirements, two critical 
questions naturally arise in regard to improving theoretical predictions 
in galaxy clustering. 1) Various effects contribute to galaxy clustering, 
such as the redshift-space distortion, the gravitational lensing, and so on. 
What is the exhaustive list of all the contributions to galaxy clustering?
We need a complete description of all the effects in galaxy clustering
to control systematics in theoretical modeling.
2) On large scales, perturbations such as
the matter density fluctuation are gauge-dependent, as there is no unique
choice of hypersurface of simultaneity throughout the entire Universe. 
The standard description of galaxy clustering is ill-posed to address this
issue, i.e., which gauge condition needs to be chosen to describe the observed 
galaxy clustering and why? 

These questions can be tackled, as we note that
the standard description of galaxy clustering is Newtonian, in which
the speed of light is infinite and the gravity is felt instantaneously
across the Universe. The light we measure in galaxy surveys, of course,
propagates at the finite speed through the Universe and is affected by
the inhomogeneity and the curvature of the Universe. Therefore, we need
proper general relativistic treatments to relate the observables we measure
from the light to the physical quantities of source galaxies and the 
inhomogeneities that affect the photon propagation.
This goal can be readily achieved by tracing back the photon path 
given the observed redshift and the angular position of the source galaxies,
and the full relativistic formula of galaxy clustering is constructed from
the observable quantities, providing the relation to the inhomogeneities
and the source galaxy population \cite{YOFIZA09,YOO10}.  In this way,
the relativistic description of galaxy clustering naturally answers the two
key questions, since observable quantities are gauge-invariant and receive
all the contributions without any theoretical prejudice.
The relativistic formula is independently developed
\cite{BODU11,CHLE11,JESCHI12}, and many interesting applications
are investigated (e.g., 
\cite{MCDON09,BASEET11,BRCRET12,SCJE12a,JESC12,BEMAET12,LOMERI12,HABOCH12,
YOHAET12,MAZHET13,LOYOKO13,YODE13,YOSE13a,BOHUGA14}).

Regarding the detectability of the relativistic effect in galaxy clustering,
it was shown \cite{YOFIZA09} that the relativistic effect
can be measured in the angular power spectrum and
the systematic errors are larger than the cosmic variance on large scales. 
Furthermore,
using the multi-tracer technique \cite{SELJA09} to eliminate the cosmic 
variance limit on large scales, it was shown \cite{YOHAET12,LOYOKO13} that 
the galaxy power spectrum can be used to measure the relativistic effect 
with great significance in upcoming surveys and can be utilized to discriminate
alternative theories of modified gravity against general relativity
on large scales. The relativistic effect in galaxy clustering becomes 
dominant on large scales,  in which modified gravity or dark energy models
deviate from general relativity and the information about the inflationary
epoch remains intact. Therefore, it is crucial to have a proper
relativistic description to avoid misinterpretation of large-scale
measurements.

The purpose of this work is to provide a pedagogical derivation of the
relativistic description of galaxy clustering. We begin by providing the
relation of the photon path to the observed redshift and the angular position
of source galaxies (section~\ref{sec:obs}). Using the observable quantities,
we construct galaxy clustering observables (section~\ref{sec:greingc})
and compute galaxy two-point statistics (section~\ref{sec:probe}).
We conclude with a discussion of further applications 
(section~\ref{sec:future}).
Throughout the paper, we use the Latin indices for the
spacetime component and the Greek indices for the spatial component, and
we set the speed of light $c\equiv1$. Symbols used in this paper are
summarized in Table~\ref{tab:tab}.

\section{Observed Angular Position and Redshift of Sources}
\label{sec:obs}

\subsection{Metric Perturbations and Gauge Transformation}
\label{ssec:metric}
The background Friedmann-Lema{\^\i}tre-Robertson-Walker (FLRW) universe 
is described by a spatially homogeneous and isotropic
\beeq
\label{eq:bgmetric}
ds^2=g_{ab}~dx^adx^b=-a^2(\tau)d\tau^2
+a^2(\tau)\gbar_{\alpha\beta}dx^\alpha dx^\beta~,
\eneq
where $\tau$ is the conformal time, $a(\tau)$ is the expansion scale factor,
and $\gbar_{\alpha\beta}$ is the 3-spatial 
metric tensor with a constant spatial curvature~$K$.
The real universe is inhomogeneous, and small deviations from the background
metric are represented by
\bear
\label{eq:abc}
\delta g_{00}&\equiv&-2~a^2\AA\equiv-2~a^2\alpha~, \qquad
\delta g_{0\alpha}\equiv-a^2\BB_\alpha\equiv
-a^2(\beta_{,\alpha}+B_\alpha) ~, \\
\delta g_{\alpha\beta}&\equiv&2~a^2\CC_{\alpha\beta}\equiv
2~a^2\left(\varphi~\gbar_{\alpha\beta}+\gamma_{,\alpha|\beta}
+{1\over2}C_{\alpha|\beta}+{1\over2}C_{\beta|\alpha}
+C_{\alpha\beta}\right)~, \nonumber
\enar
where the vertical bar is the covariant derivative with respect to
spatial metric $\gbar_{\alpha\beta}$. Perturbations are further decomposed
into scalar ($\alpha,\beta,\varphi,\gamma)$, vector ($B_\alpha,C_\alpha$) 
and tensors ($C_{\alpha\beta}$), which are readily distinguishable
by their spatial indicies. 
The fluid quantities are described by the energy-momentum tensor
\beeq
\label{eq:emo}
T_{ab}=\rho \UU_a\UU_b+p(g_{ab}+\UU_a\UU_b)+q_a\UU_b+q_b\UU_a+\pi_{ab}~,\quad
\UU^aq_a=0~,\quad \UU^a\pi_{ab}=0~,\quad \pi^a_a=0~,
\eneq
where $\rho$ is the energy density, $p$ is the isotropic pressure, $q_a$ is
the energy flux, and $\pi_{ab}$ is the anisotropic pressure. 
These fluid quantities
are measured by the the observer moving with timelike four velocity
($u^au_a=-1$)
\beeq
\label{eq:uuu}
\UU^a\equiv{1\over a}\left(1-\AA,~\VV^\alpha\right)~,\qquad
\VV^\alpha\equiv-\SVV^{,\alpha}+\VVV^\alpha~,
\eneq
where we decomposed the four velocity into scalar ($\SVV$) and vector
($\VVV^\alpha$). We also define a scalar velocity $\sv\equiv\SVV+\beta$.

The general covariance is a symmetry in relativistic theory, 
and any coordinate system can be used to describe the physical system, 
providing ample degree of freedom at hand.
However, in a cosmological framework, a coordinate transformation accompanies
a change in the correspondence of the inhomogeneous universe to
the fictitious background universe, known as the gauge transformation.
Therefore, it is important to check that theory under consideration is
gauge-invariant. For the most general coordinate transformation
\beeq
\tilde x^a=x^a+\xi^a~,\quad
\xi^a=(T,\mathcal{L}^\alpha)~,\quad
\mathcal{L}^\alpha\equiv L^{,\alpha}+L^\alpha~,
\label{eq:gt}
\eneq
the scalar quantities transform as
\beeq
\tilde\alpha=\alpha-T'-\HH T    ~,\quad
\tilde\beta=\beta-T+L'         ~,\quad
\tilde\varphi=\varphi-\HH T    ~, \quad
\tilde\gamma=\gamma-L     ~, \quad 
\tilde\SVV=\SVV-L'~,\quad
\tilde\sv=\sv-T~,
\eneq
and the vector metric perturbations transform as
\beeq
\tilde B_\alpha=B_\alpha+L'_\alpha    ~, \quad
\tilde C_\alpha=C_\alpha-L_\alpha    ~,\quad
\tilde\VVV_\alpha=\VVV_\alpha+L'_\alpha~,\quad
\eneq
where the prime is the derivative with respect to the conformal time
and the conformal Hubble parameter is $\HH=a'/a=aH$.
Since tensor harmonics are independent of tensors
that can be constructed from coordinate transformations,
tensor-type perturbations ($C_{\alpha\beta},\pi_{\alpha\beta}$)
remain unchanged under the gauge transformation in Eq.~(\ref{eq:gt}).
Similarly, the gauge-transformation properties of the fluid quantities
can be derived, and in particular the matter density fluctuation 
$\delta=\delta\rho_m/\bar\rho_m$ 
transforms as $\widetilde{\delta}=\delta+3\HH T$.

Based on the above gauge transformation properties, we can construct 
linear-order gauge-invariant quantities. The scalar gauge-invariant variables
are
\beeq
\label{eq:sgi}
\ax=\alpha-{1\over a}~\chi'~,\quad 
\px=\varphi-H\chi~,\quad \vx=\sv-{1\over a}\chi~,\quad
\dv=\delta+3\HH\sv~,
\eneq
where $\chi=a~(\beta+\gamma')$ is the scalar shear of the normal observer 
($n_\alpha=0$) and
it is spatially invariant, transforming as  
$\tilde\chi=\chi-aT$~. The notation for scalar
gauge-invariant variables is set up, such that $\dv$, for example,
corresponds to the matter density fluctuation~$\delta$ in the comoving 
gauge ($\sv=0$) and $\vx$ corresponds to the scalar velocity~$\sv$
in the zero-shear gauge ($\chi=0$). Similarly, 
$\sv_\delta=\dv/3\HH=\sv+\delta/3\HH$ 
would corresponds to the scalar velocity~$\sv$
in the uniform density gauge ($\delta=0$), and many other gauge-invariant
variables can be constructed in this way the gauge correspondence is 
explicit \cite{HWNO01}.

The vector gauge-invariant variables are
\beeq
\label{eq:vgiv}
\pv_\alpha=B_\alpha+C'_\alpha~,\qquad \gv_\alpha=\VVV_\alpha-B_\alpha ~.
\eneq
These gauge-invariant variables ($\ax,\px,\vx,\pv_\alpha,gv_\alpha$)
correspond to $\Phi_A$, $\Phi_H$,
$v^{(0)}_s$, $\Psi$, and $v_c$ in Bardeen's notation \cite{BARDE80}.
For future use, we define a gauge-invariant velocity quantity,
\beeq
\VGI_\alpha=-{\vx}_{,\alpha}+\gv_\alpha~,
\eneq
which encompasses the scalar and vector gauge-invariant variables.

Any physical quantities or observable quantities should be gauge-invariant,
i.e., the choice of gauge condition for the perturbation should be explicit.
One can achieve this goal by choosing a gauge condition, before any
calculations are performed.
However, all the quantities in this case become automatically
gauge-invariant, depriving the way to verify if the quantities of interest
are genuinely gauge-invariant. With fully general metric representation,
the verification is explicit in the calculations \cite{YOO10}.

\subsection{Observables in the Observer Rest Frame}
\label{ssec:obs}
Galaxy positions in a redshift survey are identified
by measuring photons from the sources by the observer in the rest frame.
The photon propagation direction is set orthogonal to the
hypersurface defined by constant phase 
$\vartheta=\bdi{k}\cdot\bdi{x}-\omega t$. In the observer rest frame, 
the components of the photon wavevector can be written as
\beeq
k^a_L=\eta^{ab}\vartheta_{,b}
=\left(\omega~,~\bdi{k}\right)=2\pi\nu~(1~,~-\nhat)~,
\label{eq:Lwave}
\eneq
where the local metric is Minkowsky~$\eta_{ab}$,
the angular frequency is $\omega=2\pi\nu$, and the amplitude of a
photon wavevector is $|\bdi{k}|=\omega=2\pi/\lambda$ ($\lambda\nu=1$). 
The subscript~$L$ is used to emphasize that the components
are written in the observer rest frame, or the local Lorentz frame.
The observed angular position of the source galaxy
is then determined by a unit directional vector 
$\nhat=(\ttt,\pp)$ for photon propagation in the observer rest frame,
\beeq
\nhat=-{\bdi{k}\over|\bdi{k}|}=(\sin\ttt\cos\pp,~\sin\ttt\sin\pp,~\cos\ttt)~,
\label{eq:oang}
\eneq
and the photon frequency measured by the observer is 
\beeq
\omega=2\pi\nu=-\eta_{ab}u^a_L k^b_L~,
\label{eq:freq}
\eneq
where $u^a_L=(1,0,0,0)$ in the observer rest frame. The observed
redshift of the source galaxy is then determined by the ratio of the observed 
photon wavelength $\lambda_L=1/\nu$ to the wavelength 
$\lambda_s$ we would measure
in the rest frame of the source galaxy
\beeq
1+\zz\equiv{\lambda_L\over\lambda_s}~,
\label{eq:zrat}
\eneq
where we omitted the subscript ``obs'' for the observed 
redshift~$\zz$.\footnote{We always use~$\zz$ to refer to the observed 
redshift.}
A prominent Ly$\alpha$ line $\lambda_s=121.6$~nm, for example, is often used 
to measure the redshift of source galaxies. In this way, the observed angular
position and the redshift of the source are expressed in terms of physical
quantities.

In order to compute the photon wavevector in a FRW coordinate (as opposed
to the local Lorentz frame), we first construct an orthonormal basis.
The observer is moving with a time-like four velocity $u^a$, which defines the
proper time-direction $[e_t]^a\equiv u^a$ in the observer rest frame.
Spatial hypersurface orthogonal to~$u^a$ can be described by three spacelike 
four vectors $[e_i]^a$ ($1=[e_i]^a[e_i]_a$, $i=x,y,z$). 
These four vectors, called tetrads, form an orthonormal basis in the observer
rest frame. Utilizing the orthonormality condition
($g_{ab}^L=\eta_{ab}=g_{ab}[e_c]^a[e_d]^b$, $c,d=t,x,y,z$),
the tetrads in an inhomogeneous universe with the
metric tensor in Eqs.~(\ref{eq:bgmetric}) and~(\ref{eq:abc}) 
can be constructed as
\beeq
\label{eq:tetrad}
[e_t]^a=u^a~,\qquad 
[e_i]^a={1\over a}\left[\VV_i-\BB_i~,~\delta^\alpha_i-\CC^\alpha_i\right]~,
\eneq
where the tetrad index can be raised or lowered by~$\eta_{ab}$, while
the FRW index is raised or lowered by $g_{ab}$.
The photon wavevector in a FRW coordinate can be derived by transforming
Eq.~(\ref{eq:Lwave}) in the local Lorentz frame as
\beeq
\label{eq:Fwave}
k^0={2\pi\nu\over a}\left[1-\AA-n^i(\VV_i-\BB_i)\right]~,\quad
k^\alpha={2\pi\nu\over a}\left[-n^\alpha+\VV^\alpha+n^i\CC_i^\alpha\right]~,
\eneq
where $n^i$ is the $i$-th
spatial component of the unit directional vector $\nhat$
in a local Lorentz frame, other perturbation quantities are those
in a FRW frame, and
the repeated indices indicate the summation over the spatial
components. It is noted that the photon wavevectors 
in Eqs.~(\ref{eq:Lwave}) and~(\ref{eq:Fwave}) are different, and the 
components in a FRW frame are affected by the observer velocity and the
gravitational potential.
Naturally, a unit directional vector $-k^\alpha/|k^\alpha|$ 
in a FRW frame cannot be
used to describe the observed angular position $\nhat$ in the observer
rest frame. However, the photon frequency measured by the observer 
is a Lorentz scalar, i.e., same in both frames
\beeq
-g_{ab}u^ak^b=-\eta_{ab}u_L^a k^b_L=\omega=2\pi\nu~.
\label{eq:nu}
\eneq
Likewise, the observed redshift in Eq.~(\ref{eq:zrat}) is a Lorentz scalar.

\subsection{Photon Wavevector and Conformal Transformation}
\label{ssec:pho}
We parametrize the photon path $x^a(\oo)$ with an affine parameter~$\oo$, 
and its propagation direction is then $k^a(\oo)=dx^a/d\oo$ in 
Eq.~(\ref{eq:Fwave}), 
subject to the null condition $k^ak_a=0$ and the geodesic equation
$k^a_{\;\;;b}k^b=0$. Since null geodesic is conformally
invariant, we further simplify the photon propagation equations by using
a conformal transformation $g_{ab}\rightarrow a^2\CG_{ab}$.
The null geodesic $x^a(\oo)$ described by the conformally transformed
wavevector $\CK^a$ remains unaffected under the conformal 
transformation, while its affine parameter is transformed to
another affine parameter~$\cc$ (see, e.g., \cite{WALD84})
\beeq
\label{eq:af}
{d\oo\over d\cc}=\NC a^2~,
\eneq
where the proportionality constant~$\NC$ represents 
additional degree of freedom from the conformal transformation.
The conformally transformed photon wavevector 
$\CK^a=\NC a^2k^a$ can be explicitly written as
\beeq
\CK^0=2\pi\NC \nu a\left[1-\AA-n^i(\VV_i-\BB_i)\right]~,\qquad
\CK^\alpha=2\pi\NC \nu a\left[-n^\alpha+\VV^\alpha+n^i\CC_i^\alpha\right]~.
\eneq
It proves convenient to choose the normalization constant 
$2\pi\NC a\nu\equiv1$ at the observer position $x^a(\cc_o)$, and
this normalization condition implies that the conformally transformed
wavevector can be written as
\beeq
\CK^a\equiv(1+\dnu~,-n^\alpha-\dea^\alpha)~,
\label{eq:Cwave}
\eneq
where the observed angle $n^\alpha$ measured in the observer rest frame 
is constant. The product
$2\pi\NC\nu a$ is unity at the observer position to all orders in perturbation,
and unity everywhere in a homogeneous universe. However, in an inhomogeneous
universe
it varies along the photon path as fictitious observers measure~$\nu$
in Eq.~(\ref{eq:nu}) and the scale factor~$a$ changes at each point.

For the coordinate transformation in Eq.~(\ref{eq:gt}), 
these perturbations to the wavevector transform as
\beeq
\label{eq:pertr}
\widetilde{\dnu}=\dnu+2\HH T+{d\over d\cc}~T~,  \qquad
\widetilde{\dea}^\alpha=\dea^\alpha+2\HH Tn^\alpha-{d\over d\cc}
\mathcal{L}^\alpha~,
\eneq
and we can define two gauge-invariant variables for these perturbations
\beeq
\label{eq:dnugi}
\dnug=\dnu+2H\chi+{d\over d\cc}\left({\chi\over a}\right)~,   \qquad
\deag=\dea^\alpha+2H\chi ~n^\alpha-{d\over d\cc}~\CCG^\alpha~,
\eneq
where we used the background photon path 
$d/d\cc=\partial_\tau-n^\alpha\partial_\alpha$ and
$\CCG^\alpha=\gamma^{,\alpha}+C^\alpha$ is a pure gauge term,
transforming as $\tilde\CCG^\alpha=\CCG^\alpha-\mathcal{L}^\alpha$.
The pure gauge term will be absent in observable quantities below.

\subsection{Gauge-Invariant Geodesic Equation}
\label{ssec:gige}
Having established the gauge-transformation properties of the photon 
wavevector, we now derive the photon geodesic equation.
First, the null condition of the photon wavevector is
\beeq
0=\CK^a\CK_a=\left(n^\alpha n_\alpha-1\right)
+2\left(n^\alpha\dea_\alpha-\dnu-\AA
+\BB_\alpha n^\alpha+\CC_{\alpha\beta}n^\alpha n^\beta\right)~,
\label{eq:null}
\eneq
and the background relation is trivially satisfied by the construction 
of the unit directional vector $n^\alpha$.
In terms of the gauge-invariant variables, the null condition implies
\beeq
n_\alpha ~\deag=\dnug+\ax-\px-\pv_\alpha ~n^\alpha
-C_{\alpha\beta}~n^\alpha n^\beta~.
\label{eq:GInull}
\eneq
Similarly for the geodesic equation ($k^bk^a_{~;b}=0$), the background
relation is trivially removed, and it yields the propagation equation for
the perturbation ($\dnu,\dea^\alpha$).
The temporal and spatial components of the geodesic equation are
\beeq
\label{eq:geod}
0=\CK^a\CK^0_{\;\; ;a}={d\over d\cc}\dnu+\GG^0~,\qquad
0=\CK^b\CK^\alpha_{\;\; ;b}=-{d\over d\cc}\dea^\alpha+\GG^\alpha~,
\eneq
where we have defined $\GG^0$ and $\GG^\alpha$
using the Christoffel symbol $\hat\Gamma^a_{bc}$ 
based on the conformally transformed metric $\CG_{ab}$ as
\bear
\GG^0&\equiv&\hat\Gamma^0_{ab}\CK^a\CK^b=\AA'-2\AA_{,\alpha}n^\alpha
+\left(\BB_{\alpha|\beta}+\CC_{\alpha\beta}'\right) n^\alpha n^\beta~\\
&=&2{d\over d\cc}\ax-(\ax-\px)'+(\pv_{\alpha|\beta}+C'_{\alpha\beta}n^\alpha
n^\beta)+2{d\over d\cc}H\chi+{d^2\over d\cc^2}\left({\chi\over a}\right)
~,\nonumber \\
\GG^\alpha&\equiv&\delta(\hat\Gamma^\alpha_{bc}\CK^b\CK^c)
=\AA^{,\alpha}-\BB^{\alpha\prime}
- \left( \BB_\beta^{\;\;|\alpha} - \BB^\alpha_{\;\;|\beta}
+ 2 \CC^{\alpha\prime}_\beta \right) n^\beta+
\left( 2 \CC^\alpha_{\beta|\gamma} - \CC_{\beta\gamma}^{\;\;\;\;|\alpha}
\right) n^\beta n^\gamma~\\
&=&\left(\ax-\px-\pv_\beta n^\beta-C_{\beta\gamma}n^\beta n^\gamma
\right)^{|\alpha}-{d\over d\cc}\left(2\px n^\alpha+\pv^\alpha+2C^\alpha_\beta
n^\beta\right)-{d\over d\cc}\left(2H\chi n^\alpha\right)+{d^2\over d\cc^2}
\CCG^\alpha~.\nonumber
\enar
Rearranging in terms of the gauge-invariant variables, we derive the
gauge-invariant geodesic equations for temporal component
\beeq
{d\over d\cc}~\left(\dnug+2~\ax\right)=(\ax-\px)'-\left(\pv_{\alpha|\beta}
+C'_{\alpha\beta}\right)n^\alpha n^\beta~,
\label{eq:GItemp}
\eneq
and for spatial component
\beeq
\label{eq:GIsp}
{d\over d\cc}\left(\deag+2~\px n^\alpha+\pv^\alpha+2~C^\alpha_\beta n^\beta
\right)
=\left(\ax-\px-\pv_\beta n^\beta-C_{\beta\gamma}n^\beta n^\gamma
\right)^{|\alpha}~.
\eneq
Fictitious gauge freedoms in $\dnu$ and $\dea^\alpha$ are
completely removed, and Eqs.~(\ref{eq:GInull}), (\ref{eq:GItemp}),
and~(\ref{eq:GIsp}) are manifestly gauge-invariant.

\subsection{Observed Source Position and Redshift}
\label{ssec:redshift}
The source galaxy position on the sky is identified by the observed
angle $(\ttt,\pp)$ in Eq.~(\ref{eq:oang}) and the observed 
redshift~$\zz$ in Eq.~(\ref{eq:zrat}).
Based on these observables, the observers infer the source position  by
using the distance-redshift relation in a homogeneous universe, i.e., 
\beeq
\label{eq:inferred}
\hat x^a_s\equiv[\bar\tau_z~,~\rbar_z \nhat]
=[\bar\tau_z~,~\rbar_z\sin\ttt\cos\pp~,~\rbar_z\sin\ttt\sin\pp~,
~\rbar_z\cos\ttt]~,
\eneq
where the source position is expressed in a rectangular coordinate,
the comoving distance to the source is
\beeq
\label{eq:bar}
\rbar_z\equiv\rbar(\zz)=\bar\tau_o-\bar\tau_z=\int_0^\zz{dz'\over H(z')}~,
\eneq
and a bar is used to indicate that these quantities are computed at the 
background level. Given a set of cosmological parameters, these quantities
are fully determined, and there are no gauge ambiguities associated with
coordinate transformations.

However, the real position $x^a_s$ of the source galaxy
is different from the inferred
source position $\hat x^a_s$, as the universe is inhomogeneous, affecting
the photon propagation, and the components of $x^a_s$ themselves are 
gauge-dependent. To represent the source position with respect 
to the inferred source position, we define the coordinate distortions
$(\dTau,\drr,\dtt,\dpp)$ by expressing the real source galaxy position as
\beeq
\label{eq:disp}
x^a_s\equiv\left[\bar\tau_z+\dTau
~,~(\rbar_z+\drr)\sin(\ttt+\dtt)\cos(\pp+\dpp)~,~(\rbar_z+\drr)
\sin(\ttt+\dtt)\sin(\pp+\dpp)~,~(\rbar_z+\drr)\cos(\ttt+\dtt)\right]~,
\eneq
and it is noted that these coordinate distortions are also gauge-dependent,
as we show below. The coordinate distortions can be computed by tracing the
photon path backward from the observer and solving for $x^a_s$. First, we
consider the source galaxy position $\bar x^a_s$ in a homogeneous universe
by integrating the photon wavevector in Eq.~(\ref{eq:Cwave}) over the 
affine parameter~$\cc$ as
\beeq
\label{eq:shomo}
\bar x^a(\cc_s)-\bar x^a(\cc_o)
=\left[\bar\tau_s-\bar\tau_o,~\bar x^\alpha_s\right]
=\left[\cc_s-\cc_o,~(\cc_o-\cc_s)n^\alpha\right]~,
\eneq
and without loss of generality we set $\bar x^\alpha(\cc_o)=\bar x^\alpha_o=0$,
and $\cc_o=0$.
The relation in Eq.~(\ref{eq:shomo}) defines the affine parameter in a
given coordinate system as
\beeq
\cc=\bar\tau-\bar\tau_o=-\rbar=-\int_0^{\zh}{dz'\over H(z')}~,
\label{eq:affine}
\eneq
but in terms of the redshift parameter $1+\zh(\tau)\equiv1/a(\tau)$ 
(or coordinate time $\tau$) of the source position in the background
(note that the observed redshift is~$\zz$).
Since it depends on the coordinate time of the source position, 
the redshift parameter transforms $\widetilde{\zh}=\zh-HT$
under the coordinate transformation in Eq.~(\ref{eq:gt}), 
and so does the affine parameter 
\beeq
\tilde\cc=\cc\left(1+\HH_oT_o\right)-\int_0^\cc d\cc'~2\HH T~,
\label{eq:afftr}
\eneq
where the integration over the affine parameter represents that the integrand
is evaluated along the photon path~$x^a_\cc$.

The coordinate distortions are useful quantities for characterizing the source
galaxy position $x^a_s$, as the observer uses the inferred source position
$\hat x^a_s$ based on the observable quantities. In the same way, it is
convenient to define the affine parameter $\cc_z$, satisfying 
Eq.~(\ref{eq:affine}) in terms of the observed redshift~$\zz$, instead
of a redshift parameter~$\zh$. The affine parameter at the 
source position~$x^a_s$ is parametrized as
\beeq
\cc_s\equiv\cc_z+\Dcc_s~,
\eneq
and the (conformal) time coordinate of the source galaxy position
can be rephrased as\footnote{At a given affine parameter~$\cc$, the
position $x^a_\cc$ along the photon path can be split into the mean and
the perturbation: $x^a_\cc=\bar x^a_\cc+\delta x^a_\cc$, where $\bar x^a_\cc$
represents the position in a homogeneous universe at the same~$\cc$ and
$\delta x^a_\cc$ represents the residual perturbation in $x^a_\cc$.
So it is noted that $\tau_s=\bar\tau_s+\delta\tau_s=\bar\tau_z+\dTau$.}
\beeq
\label{eq:DT}
\tau_s\equiv\tau(\cc_s)=\bar\tau(\cc_z+\Dcc_s)+\dT(\cc_z+\Dcc_s)=
\bar\tau_z+\Dcc_s+\dT_z~,
\eneq
where the subscript~$\zz$ indicates that quantities are evaluated at the 
observed redshift (or the affine parameter~$\cc_z$). Since 
the observed redshift is related to $\bar\tau_z$ as
\beeq
\label{eq:Obsz}
1+\zz={1\over a(\bar\tau_\zz)}={\lambda_L\over\lambda_s}
={\left(k^au_a\right)_s\over\left(k^au_a\right)_o}
\equiv{1+\dz\over a(\tau_s)}~,
\eneq
substituting Eq.~(\ref{eq:DT}) yields that the distortion $\DT$ in
time coordinate is related to the distortion $\dz$ in the observed redshift
\beeq
\DT=\Dcc_s+\dT_z={\dz\over\HH_z}~.
\eneq
Finally, we evaluate the photon frequency along the photon path $x^a_\cc$
by using Eqs.~(\ref{eq:freq}) and~(\ref{eq:nu})
\beeq
\label{eq:cnu}
2\pi\nu=-k^au_a=-{\CK^a\CU_a\over\NC a}
={1\over\NC a}\left[1+\dnu+\AA+\left(\VV_\alpha-\BB_\alpha\right)n^\alpha
\right]
={1\over\NC a}\left[1+\dnug+\ax+\VGI_\alpha n^\alpha-H\chi\right]~,
\eneq
and the distortion in the observed redshift is
\bear
\label{eq:zobs}
\dz&=&\HH_o\dT_o+\bigg[\dnug+\ax+\VGI-H\chi\bigg]^s_o \\
&=&-H\chi+(H_o\chi_o+\HH_o\dT_o)+\bigg[V-\ax\bigg]^z_o 
-\int_0^{\rbar_z}d\rbar\bigg[(\ax-\px)'-(\pv_{\alpha|\beta}
+C'_{\alpha\beta})~n^\alpha n^\beta\bigg]~, \nonumber
\enar
where $\rbar_z$ is the comoving line-of-sight distance to the source,
$\VGI=\VGI_\alpha n^\alpha$ is the line-of-sight velocity, and
$\tau(\cc_o)=\bar\tau_o+\dT_o$.
We also used Eq.~(\ref{eq:GItemp}) for integration.
Spurious spatial gauge freedom in~$\dz$ is removed
and its temporal gauge dependence $\widetilde{\dz}=\dz+\HH T$ leaves
the observed redshift~$\zz$ explicitly gauge-invariant.
We define a gauge-invariant variable for the lapse in the observed redshift
as $\dzg=\dz+H\chi$~.

\section{Relativistic Description of Galaxy Clustering}
\label{sec:greingc}

\subsection{Geometric Distortions in Photon Path}
\label{ssec:path}
Now we derive the geometric distortions $(\drr,\dtt,\dpp)$
in Eq.~(\ref{eq:disp}). Since we
clarified the relation between the photon wavevector and the observable
quantities, the photon wavevector can be integrated over the affine 
parameter to obtain the source galaxy
position and express it in terms of the
observable quantities. The integration over the affine parameter can be
subsequently converted into the integration over the mean photon path 
$d\rbar=-d\cc$
in Eq.~(\ref{eq:affine}), as we are interested in the linear order effect.
Therefore, noting that the affine parameter describing the source galaxy
position is
$\cc_s=\cc_z+\Dcc_s$, we first integrate the geodesic equation 
in Eq.~(\ref{eq:geod}) to relate
the perturbations to the photon wavevector with metric perturbations as
\bear
\dnu\bigg|_o^\zz&=&-2(\AA_z-\AA_o)-\int_0^{\rbar_z} d\rbar~
\left[\AA'-\left(\BB_{\alpha|\beta}+\CC_{\alpha\beta}'\right)n^\alpha 
n^\beta\right]\\
&=&-2\ax\bigg|^\zz_o-\int_0^{\rbar_z}d\rbar
\bigg[(\ax-\px)'-\left(\pv_{\alpha|\beta}
+C'_{\alpha\beta}\right)n^\alpha n^\beta\bigg]
-2H\chi\bigg|^\zz_o-{d\over d\cc}\left({\chi\over a}\right)\bigg|^\zz_o
~,\nonumber \\
\dea^\alpha\bigg|_o^\zz&=&
-\bigg[\BB^\alpha+2\CC^\alpha_\beta n^\beta\bigg]^\zz_o-\int_0^{\rbar_z}
d\rbar~\left(\AA-\BB_\beta n^\beta-\CC_{\beta\gamma}n^\beta n^\gamma\right)
^{|\alpha}\\
&=&-\bigg[2~\px n^\alpha+\pv^\alpha+2~C^\alpha_\beta n^\beta\bigg]^\zz_o
-\int_0^{\rbar_z}d\rbar\left(\ax-\px-\pv_\beta n^\beta
-C_{\beta\gamma}n^\beta n^\gamma\right)^{|\alpha}
-2H\chi n^\alpha\bigg|^\zz_o+{d\over d\cc}\CCG^\alpha\bigg|^\zz_o~,\nonumber
\enar
and then integrate the photon wavevector to obtain the source position 
\bear
\label{eq:src}
x^a_s&=&\left[\bar\tau_z+\dT_o+\Dcc_s-\int_0^{\rbar_z}d\rbar~\dnu~,
~\rbar_z n^\alpha-\Dcc_s n^\alpha +\int_0^{\rbar_z}d\rbar~\dea^\alpha\right] \\
&=&\left[\bar\tau_z+\dT_o+\cc_z\dnu_o+\Dcc_s
-\int_0^{\rbar_z}d\rbar~(\rbar_z-\rbar)
~\GG^0~,~\rbar_z n^\alpha+\rbar_z\dea^\alpha_o-\Dcc_s n^\alpha
-\int_0^{\rbar_z}d\rbar~(\rbar_z-\rbar)~\GG^\alpha\right]~. \nonumber
\enar
where quantities ($\dnu_o,\dea^\alpha_o$) at the observer position.
Constructing two unit directional vectors based on the observed angle
\beeq
\thatv={\partial\over\partial\ttt}\nhat
=(\cos\ttt\cos\pp~,\cos\ttt\sin\pp~,-\sin\ttt)~,\qquad
\phatv={1\over\sin\ttt}{\partial\over\partial\pp}\nhat=
(-\sin\pp~,\cos\pp~,0)~,
\eneq
the geometric distortions of the source galaxy position can be explicitly
computed in terms of metric perturbations as
\bear
\label{eq:drr}
\drr&=&n_\alpha x^\alpha_s-\rbar_z=
-\Dcc_s+\int_0^{\rbar_z}d\rbar~n_\alpha\dea^\alpha=
\dT_o-{\dz\over\HH_z}+\int_0^{\rbar_z}d\rbar~\left(
\AA-\BB_\alpha n^\alpha-\CC_{\alpha\beta}n^\alpha n^\beta \right)~\\
&=&\left(\chi_o+\dT_o\right)-{\dzg\over\HH_z}+\int_0^{\rbar_z}d\rbar
\bigg[(\ax-\px)'-\left(\pv_{\alpha|\beta}
+C'_{\alpha\beta}\right)n^\alpha n^\beta\bigg]-n_\alpha\CCG^\alpha\bigg|^\zz_o
~,\nonumber \\
\label{eq:dtt}
\rbar_z\dtt&=&\ttt_\alpha x^\alpha_s=
\rbar_z\theta_\alpha\dea^\alpha_o
-\int_0^{\rbar_z}d\rbar~(\rbar_z-\rbar)~\theta_\alpha\GG^\alpha\\
&=&\rbar_z\theta_\alpha\left(\dea^\alpha+\BB^\alpha+2~\CC^\alpha_\beta
n^\beta\right)_o-\int_0^{\rbar_z}d\rbar~\left[\theta_\alpha
\left(\BB^\alpha+2~\CC^\alpha_\beta n^\beta\right)+\left({\rbar_z-\rbar\over
\rbar}\right){\partial\over\partial\ttt}
\left(\AA-\BB_\alpha n^\alpha-\CC_{\alpha\beta}n^\alpha n^\beta \right)
\right]~\nonumber\\
&=&\rbar_z\theta_\alpha\left(\deag+\pv^\alpha+2~C^\alpha_\beta
e^\beta\right)_o-\theta_\alpha\CCG^\alpha\bigg|^\zz_o \nonumber \\
&&-\int_0^{\rbar_z}d\rbar~\left[\theta_\alpha
\left(\pv^\alpha+2~C^\alpha_\beta n^\beta\right)+\left({\rbar_z-\rbar\over
\rbar}\right){\partial\over\partial\ttt}
\left(\ax-\px-\pv_\alpha n^\alpha-C_{\alpha\beta}n^\alpha n^\beta \right)
\right]~,\nonumber 
\enar
where we used Eq.~(\ref{eq:GInull}) for manipulating $\drr$. 
The azimuthal distortion $\rbar_z\sin\ttt\dpp$ is similar to $\rbar_z\dtt$.
It is apparent
that these geometric distortions are gauge-dependent quantities. Physically,
the radial and angular distortions arise due to the metric perturbations
along the photon path and the identification of the source at the observed
redshift.

\subsection{Lensing Magnification and Luminosity Distance}
\label{ssec:kappa}
Here we derive two quantities associated with angular distortions of the
source galaxy 
position on the sky. The first quantity, known as the gravitational
lensing convergence~$\kappa$, describes the change in the
solid angle as part of the distortion in the physical volume. As such, 
the convergence itself is not directly associated with observable quantities.
The second quantity $\dL(\zz)$ describes the luminosity distance of 
a standard candle with known luminosity
in the rest frame at the observed redshift. Naturally, the luminosity distance
is an observable quantity, and its perturbation~$\ddL$ is related to the
gravitational lensing convergence~$\kappa$.

We first compute the gravitational lensing convergence~$\kappa$.
In galaxy clustering, we only need the change in the solid angle between
the observed $(\ttt,\pp)$ and the (unobserved) source $(\ttt+\dtt,\pp+\dpp)$,
and
the ratio of the solid angles is the Jacobian of the angular transformation
or the determinant of the deformation matrix:
\beeq
\label{eq:kappadef}
\left|{\partial(\ttt+\dtt,\pp+\dpp)\over \partial(\ttt,\pp)}\right|
={\sin(\ttt+\dtt)\over \sin\ttt}
\left[1+{\partial\over\partial\ttt}\dtt+{\partial\over\partial\pp}\dpp
\right]=1+\left(\cot\ttt+{\partial\over\partial\ttt}
\right)\dtt+{\partial\over\partial\pp}\dpp\equiv 1-2\kappa~,
\eneq
where we computed the Jacobian only to the linear order in perturbation.
At this order, the distortion in the solid angle is completely described
by the isotropic expansion in angle (gravitational lensing convergence),
and the angular shear and rotation come at higher order in perturbations.

Using Eq~(\ref{eq:dtt}), the gravitational lensing convergence can be derived
as
\bear
\label{eq:kappa}
\kappa&=&n_\alpha(\deag+\pv^\alpha+2C^\alpha_\beta n^\beta)_o
-\int_0^{\rbar_z}d\rbar~ {n_\alpha(\pv^\alpha+2C^\alpha_\beta n^\beta)\over
\rbar_z}+\int_0^{\rbar_z}d\rbar~{1\over2\rbar_z}\hat\nabla_\alpha
(\pv^\alpha+2C^\alpha_\beta e^\beta) \nonumber \\
&+&\int_0^{\rbar_z}d\rbar\left({\rbar_z-\rbar\over 2~\rbar \rbar_z}\right)
\hat\nabla^2
\bigg(\ax-\px-\pv_\alpha n^\alpha-C_{\alpha\beta}n^\alpha n^\beta\bigg)
-{n_\alpha\CCG^\alpha\over\rbar_z}\bigg|^\zz_o
+{1\over2\rbar_z}\hat\nabla_\alpha\CCG^\alpha~, 
\enar
where $\hat\nabla$ is the angular gradient operator. The gravitational lensing
convergence is gauge-dependent, as it is expressed in relation of the observed
angular positions to the unobservable source position.

Next, we compute the fluctuation $\ddL$
in the luminosity distance. The observed
flux~$f_\up{obs}$ of a source galaxy at the observed redshift~$\zz$ is used 
to infer its luminosity $\hat L=4\pi\bar D_L^2(\zz)f_\up{obs}$, 
where the luminosity distance
in a homogeneous universe is $\bar D_L(\zz)=(1+\zz)\rbar_z$. However,
the physical luminosity $L_\up{phy}\equiv4\pi\dL^2(\zz)f_\up{obs}$
of the source is different from the inferred luminosity, 
as the source galaxy is not at the inferred distance and the photon propagation
is affected by fluctuations along the path. We define the dimensionless
fluctuation in the luminosity distance $\dL(\zz)=\bar D_L(\zz)(1+\ddL)$.
As expressed in terms of
observable quantities, the fluctuation $\ddL$ is an gauge-invariant observable
quantity, as we prove below.

Since the luminosity distance is related to the angular diameter distance
$\dA(\zz)=\dL(\zz)/(1+\zz)^2$, the fluctuation in the
angular diameter distance is identical to the fluctuation in the luminosity
distance. Thus, we compute the fluctuation in the angular diameter 
distance by using the geometric distortions we already computed. 
In the source rest frame, consider a unit area $dA_\up{phy}$ 
that is perpendicular to the observed photon vector $\NN^a$ parallelly 
transported along the photon path to the source position. This unit area
would appear subtended by a solid angle 
$d\Omega=\sin\ttt d\ttt d\pp$ measured by the observer, and it is 
related to the angular diameter distance as \cite{JESCHI12}
\beeq
dA_\up{phy}=\dA^2(\zz)d\Omega=\sqrt{-g}~\varepsilon_{dabc}\UU^d_s\NN^a_s
{\partial x^b_s\over\partial\ttt}{\partial x^c_s\over\partial\pp}~d\ttt~d\pp ~,
\label{eq:dA}
\eneq
where $\sqrt{-g}\equiv a^4(1+\dg)$ is the metric determinant,
$\varepsilon_{abcd}$ is a Levi-Civita symbol,
the source position is $x^a_s$ in Eq.~(\ref{eq:src}), 
and the observed photon vector $\NN^a=k^a/(k^b\UU_b)+\UU^a$ in a FRW frame.
The covariant expression in Eq.~(\ref{eq:dA})
represents a simple mapping of the solid angle
in the observer rest frame to the physical area in the source rest frame
defined by the four velocity of the source and perpendicular to the photon
wavevector.

Removing the mean angular diameter distance $\bar D_A=\rbar_z/(1+\zz)$,
we simplify Eq.~(\ref{eq:dA}) to obtain the relation for $\ddL$
\beeq
(1+\ddL)^2=(1+\dg)(1+\dz)^2~{\varepsilon_{dabc}\over\rbar_z^2
\sin\ttt}~\CU^d_s\left(a\NN^a\right)_s
{\partial x^b_s\over\partial\ttt}{\partial x^c_s\over\partial\pp}~,
\qquad
\dg=\AA+\CC^\alpha_\alpha~,\qquad \CU^a=a\UU^a~.
\eneq
Expanding the equation to the linear order in perturbation,
the fluctuation in the luminosity distance is derived as
\beeq
\label{eq:ddl}
\ddL=\dz-\kappa+{\drr\over\rbar_z}
+{1\over2}\left(\CC^\alpha_\alpha-\CC_{\alpha\beta}e^\alpha e^\beta\right)
=\dzg-\kag+{\drr_\chi\over \rbar_z}+\px-{1\over2}C_{\alpha\beta}
n^\alpha n^\beta~,
\eneq
where we defined two-gauge invariant variables
\beeq
\label{eq:rg}
\drr_\chi=\drr+n_\alpha\CCG^\alpha\bigg|^\zz_o~,\qquad 
\kag=\kappa+{n_\alpha\CCG^\alpha\over\rbar_z}\bigg|^\zz_o
-{1\over2\rbar_z}\hat\nabla_\alpha\CCG^\alpha~
\eneq
by removing the gauge-dependent terms in Eqs.~(\ref{eq:drr})
and~(\ref{eq:kappa}), respectively.
Written in terms of gauge-invariant variables, we explicitly verify that
the fluctuation $\ddL$ is a gauge-invariant observable, and 
Eq.~(\ref{eq:ddl}) recovers the expressions for the luminosity distance 
computed in the 
conformal Newtonian gauge \cite{BODUGA06} and in the synchronous gauge
\cite{JESCHI12}.
The observed flux is affected,
not only by the change~$\kappa$ in the observed solid angle, but also affected
by the change in the radial direction set by the observed redshift.

\subsection{Physical Volume Occupied by Sources 
in 4D Spacetime}
\label{ssec:dV}
Extending the previous calculation of a unit area in the source rest frame,
we now compute the physical 3D volume occupied by the source galaxy at
$x^a_s$ in 4D spacetime that appears to the observer within the small
interval $d\zz$ of the observed redshift and the observed solid angle
$d\Omega$ \cite{WEINB72,YOFIZA09,YOO10}:
\beeq
\label{eq:dV}
dV_\up{phy}=\sqrt{-g}~\varepsilon_{dabc}~\UU^d_s~
{\partial x^a_s\over\partial\zz}
{\partial x^b_s\over\partial\ttt}{\partial x^c_s\over\partial\pp}~
d\zz~d\ttt~d\pp
\equiv d\bar V_\up{obs}(1+\dV)~, 
\qquad d\bar V_\up{obs}={\rbar_z^2~d\zz d\Omega\over H_z(1+\zz)^3}~,
\eneq
where the dimensionless volume fluctuation $\dV$ is defined with respect
to the volume $d\bar V_\up{obs}$ inferred by the observer.
Expanding the covariant expression to the linear order in perturbation, we
derive
\bear
\label{eq:ddV}
\dV&=&3~\dz+\dg+2~{\drr\over\rbar_z}-2~\kappa
+H_z{\partial\over\partial\zz}~\drr-\AA+\VV^\alpha n_\alpha \\
&=&
3~\dzg+3~\px+2~{\drr_\chi\over\rbar_z}-2~\kag+
H_z{\partial\over\partial z}\drr_\chi+\VGI+n_\alpha\pv^\alpha \nonumber\\
&=&3~\dzg+\ax+2~\px+2~{\drr_\chi\over\rbar_z} -2~\kag
-H_z{\partial\over\partial\zz}\left({\dzg\over\HH}\right)+\VGI
-C_{\alpha\beta}~n^\alpha n^\beta~. \nonumber
\enar
This equation is manifestly gauge-invariant, providing the linear-order
relativistic effect in the volume distortion. In Eq.~(\ref{eq:dV}), the
physical volume is mapped by using the three independent variables that are
the observable quantities $(\zz,\ttt,\pp)$ in the observer rest frame.
Therefore, the derivative terms in Eq.~(\ref{eq:dV}) 
(and hence in Eq.~[\ref{eq:ddV}]) are the partial derivatives with the other
observed quantities held fixed. For example, the derivative with respect 
to the observed redshift represents the change in response to the variation
in the observed redshift, which is the line-of-sight derivative 
along the past light cone. To the linear order in perturbation, it is the
background photon path, involving 
not only the spatial derivative, but also the time derivative.
As the inferred volume is $d\bar V_\up{obs}$, the volume distortion $\dV$
in the physical volume $dV_\up{phy}$ is expected to have contributions from
each component in $d\bar V_\up{obs}$. Notably, the contribution $3~\dz$ 
from the comoving factor $(1+\zz)^3$ in $d\bar V_\up{obs}$, the
contribution $2~\drr/\rbar_z$ arises from $\rbar^2_z$, the contribution
$2~\kappa$ from $d\Omega$, the contribution $H_z\partial_z\drr$ from 
the change of the radial displacement at the observed redshift, and the
remaining contribution from defining the source rest frame. As written in
terms of geometric distortions, the notation is physically transparent.

\subsection{Observed Galaxy Number Density and 
Galaxy Clustering}
\label{ssec:ng}
Given the observed redshift and angle in observation, the volume element 
$d\bar V_\up{obs}$ 
is used to infer the volume occupied by the source galaxies on the sky,
and the observed number density is obtained by counting the number of 
galaxies within the observed redshift and solid angle:
\beeq
\label{eq:obsng}
dN_g^\up{obs}(\zz,\nhat)=n_g^\up{obs}d\bar V_\up{obs}=n_gdV_\up{phy}~,\qquad
n_g^\up{obs}=n_g(1+\dV)~,
\eneq
where $n_g$ is the physical number density of source galaxies.
It is evident that the volume distortion~$\dV$ in Eq.~(\ref{eq:ddV})
always contributes to the observed galaxy number density, and its contribution
is collectively described as the volume effect \cite{MATSU00,YOO09}.

In general, the physical number density $n_g$ of source galaxies 
can be written in terms of the mean and the intrinsic fluctuation as
\beeq
\label{eq:png}
n_g=\bar n_g(t_p)\left(1+\delta_g^\up{int}\right)~,\qquad
\bar n_g(t_p)\equiv
\langle n_g\rangle_{t_p}~,\qquad \langle\delta_g^\up{int}\rangle_{t_p}=0~,
\eneq
where the mean $\bar n_g$ and the fluctuation $\delta_g^\up{int}$
are defined over a hypersurface denoted with~$t_p$. 
Galaxies are tracers of the underlying matter distribution, and the relation
between the galaxy fluctuation~$\delta_g^\up{int}$ and 
the matter fluctuation~$\delta_m$ is called galaxy bias. The galaxy bias
is known to be linear on large scales $\delta_g^\up{int}=b~\delta_m$
\cite{KAISE84}. However, in the relativistic context, this linear bias
relation is ambiguous, as the choice of gauge condition for~$\delta_m$
is unspecified. A physically meaningful choice of~$t_p$ is the proper-time
hypersurface, described by the comoving-synchronous gauge in a presureless
medium:
\beeq
\label{eq:bias}
\delta_g^\up{int}=b~\dtp~,
\eneq
where $\dtp$ is the matter density fluctuation in the 
comoving-synchronous gauge ($\dtp=\dv$). 
It is noted that the matter fluctuation in the comoving-synchronous gauge
represents one in the proper-time hypersurface {\it only} when the universe
is dominated by a presureless medium, and it becomes more subtle
beyond the linear order \cite{YOO14b}.
Galaxy formation is a local process, and its
dynamics is affected by the long wavelength modes through the change in 
the local clock that can be measured without any knowledge beyond the local
area \cite{BASEET11}. This biasing scheme is consistent with other
recent study \cite{BASEET11,BODU11,CHLE11,BRCRET12,JESCHI12,YOHAET12}
(see footnote\footnote{In
\cite{YOFIZA09,YOO10}, 
we adopted the simplest approach for biasing $n_g=F[\rho_m]$,
i.e., the physical galaxy number density is some unknown function of the
matter density at the same spacetime. While it lacks any gauge issues,
it is rather physically restrictive, as the galaxy number density evolution 
is driven only by the matter density evolution: 
$n_g=\bar n_g(\zz)(1+b~m_{\dz})$ and $e=3b$,
where $m_{\dz}=\delta_m-3~\dz$ is a gauge-invariant matter density 
fluctuation at the observed redshift. Equation~(\ref{eq:bias}) provides
a more physically motivated biasing scheme than one in \cite{YOFIZA09,YOO10}.} 
for the different biasing scheme
used in \cite{YOFIZA09,YOO10}).

The other contributions to galaxy clustering, the source effect \cite{YOO09}, 
are associated with the physical quantities of the source galaxies, but
expressed in terms of observable quantities such as the observed redshift
and flux. The mean galaxy number density is represented in the proper
time hypersurface, or the rest frame of galaxies, which is different from the
observed redshift. So when the observed number density is expressed at
the observed redshift, the physical number density is
\beeq
n_g=\bar n_g(z)\left[1-e~\dz_{t_p}\right]\left(1+\delta_g^\up{int}\right)~,
\qquad e={d\ln\bar n_g\over d\ln(1+z)}~,
\eneq
where the coefficient~$e$ is the evolution bias and the distortion in
the observed redshift is evaluated at the proper time~$t_p$. Note that
a galaxy sample with constant comoving number density would have $e=3$.
Additional contribution arises when we characterize the galaxy sample
by using its inferred luminosity as threshold. As discussed
in Sec.~\ref{ssec:kappa}, the inferred luminosity is different from the
physical luminosity, and its contribution is
\beeq
n_g=\bar n_g(\hat L)\left[1-t~\ddL\right]
\left(1+\delta_g^\up{int}\right)~,\qquad
t\equiv-2~{d\ln\bar n_g\over d\ln L}~,
\eneq
where the coefficient~$t$ describes the slope of the luminosity function.
If the differential
luminosity function $d\bar n_g(L)\propto L^{-s}$ is well approximated
 by a constant slope~$s$, we have $t=2(s-1)$. It is often the case that
the cumulative luminosity function slope $p=d\log_{10} \bar n_g(M)/dM$
is expressed in terms of magnitude
$M=\up{constant}-2.5\log_{10}(L/L_0)$, and we have $p=0.4(s-1)$ and
$t=5p$.

The observed galaxy number density is physically well-defined and is expressed
in terms of the observed redshift, angle, and the number of galaxies.
Collecting all the contributions to the observed galaxy number density, we
can relate it to various perturbation contributions as
\beeq
\label{eq:nobs}
n_g^\up{obs}
(\zz,\nhat)=\bar n_g(\zz)(1+\delta_g^\up{int})(1+\dV)(1-e~\dz_{t_p})
(1-t~\ddL)~.
\eneq
Depending on how the galaxy sample is defined in terms of other observable
quantities, additional terms from the source effect may be present 
in Eq.~(\ref{eq:nobs}). While all the perturbation contributions can be
computed, we have no means to compute the mean number density $\bar n_g(z)$ 
of the observed galaxy sample, defined in Eq.~(\ref{eq:png}). In observation,
the observed mean number density is obtained by averaging the number density
over the survey area,
\beeq
\label{eq:mmm}
\widehat{\bar n_g}(\zz)\equiv{1\over\Omega}\int_\Omega d^2\nhat~
n_g^\up{obs}(\zz,\nhat)~,
\eneq
which can introduce additional fluctuations in the estimate of the mean
galaxy number density $\bar n_g(\zz)$, since perturbations of wavelength
larger than the survey size will be absorbed in the observed mean
$\widehat{\bar n_g}$.
The observed galaxy fluctuation is then obtained by using the observed mean
number density as
\beeq
\label{eq:dobsdef}
\delta_g^\up{obs}(\zz,\nhat)\equiv 
{n_g^\up{obs}(\zz,\nhat)\over\widehat{\bar n_g}(\zz)}-1~,\qquad
n_g^\up{obs}(\zz,\nhat)=\widehat{\bar n_g}(\zz)(1+\dobs)~.
\eneq
Assuming no residual fluctuation in the mean number density, i.e.,
$\widehat{\bar n_g}(\zz)=\bar n_g(\zz)$, the galaxy fluctuation can be
written as
\bear
\label{eq:dobs}
&&\hspace{-20pt}
\delta_g^\up{obs}(\zz,\nhat)=\delta_g^\up{int}+\dV-e~\dz_{t_p}-t~\ddL\\
&&=b~\dtp-e~\dz_{t_p}-t~\ddL+
3~\dzg+\ax+2~\px+2~{\drr_\chi\over\rbar_z} -2~\kag
-H_z{\partial\over\partial\zz}\left({\dzg\over\HH}\right)+\VGI
-C_{\alpha\beta}~n^\alpha n^\beta~. \nonumber
\enar
This equation is the main result and contains all the linear-order relativistic
effects that come into play in galaxy clustering. Contributions
to galaxy clustering are physically split into the source and the volume 
effects, and they arise as the photon propagation is affected by subtle
relativistic effects between the source and observer positions. 
It is apparent in Eq.~(\ref{eq:dobs}) that the observed galaxy fluctuation
receives contributions not only from scalar perturbations, but also from
vector and tensor perturbations.

\section{Cosmological Probes: Observed Galaxy Two-Point Statistics}
\label{sec:probe}
Here we derive the galaxy two-point statistics, measurable in galaxy surveys. 
By definition, the galaxy fluctuation $\dobs(\zz,\nhat)$ 
in Eq.~(\ref{eq:dobsdef}) vanishes upon averaging over the survey area.
The galaxy two-point statistics $\la\dobs\dobs\ra$,
therefore, provides crucial ways to probe cosmology in galaxy surveys.
If the underlying distribution is Gaussian,
the two-point statistics (the power spectrum or the
correlation function) contains the complete information about the distribution,
and in practice the Universe is, to a good approximation, Gaussian
on large scales, where the relativistic effect is significant.
Higher-order statistics such as the galaxy bispectrum describes the deviation
from the Gaussianity, providing crucial clues about the initial condition 
at very early time. However, it requires the second-order relativistic
calculation \cite{YOZA14,BEMACL14,BEMACL14b,DIDUET14}, 
and here we focus on the linear-order relativistic
effect in galaxy clustering and the derivation of the observed galaxy
two-point statistics. Due to the restriction in length, we refer the 
reader to papers cited in this section for plots of two-point statistics and
their physical explanation.

\subsection{Linearized Einstein Equations and Computation
of the Observed Galaxy Fluctuation $\dobs$}
\label{ssec:eins}
Being constructed solely from observable quantities, the observed 
galaxy fluctuation $\dobs$ is gauge-invariant, as verified in 
Eq.~(\ref{eq:dobs}). Therefore, it can be evaluated with any choice of 
gauge conditions. Here we provide a simple way to compute the observed galaxy 
fluctuation, assuming general relativity.\footnote{Indeed, 
$\dobs$ in Eq.~(\ref{eq:dobs})
is derived without using Einstein equations, i.e., the general relativistic 
description of galaxy clustering is formulated in a general metric 
representation, and the only assumption was that the photons follow 
the geodesic, such that it is not restricted to Einstein's gravity, but 
applicable to other alternative theories of modified gravity 
(see, e.g., \cite{LOYOKO13}).}
The computation of galaxy two-point statistics will follow in the subsequent
sections. For simplicity, we assume that there are no vector or tensor modes
at the initial condition and ignore their contributions 
to $\dobs$ from now on.

In a flat universe with presureless medium (CDM and baryons on large
scales; $p=\pi_{ab}=0$ in Eq.~[\ref{eq:emo}]), the Einstein equations can be
arranged in terms of gauge-invariant variables 
(e.g., \cite{BARDE80,KOSA84,HWNO01}) as
\beeq
\nabla^2\px=-{3H_0^2\over2}\OM~{\dv\over a}~, \qquad \cp'=\HH\alpha_v~,\qquad
\ax=-\px~,
\eneq
and the conservation equations yields
\beeq
\label{eq:cons}
\vx'+\HH\vx=\ax~,\qquad \dv'=-3\cp'-\vx~,
\eneq
where we defined two additional scalar gauge-invariant variables,
two curvature perturbations in the comoving gauge ($\sv=0$),
$\av=\alpha-(a\sv)'/a$ and $\cp=\varphi-\HH\sv$. 
Since $\ax=\av+(a\vx)'/a$, the conservation equation for momentum
implies $\av=0$, i.e., for a flat universe with presureless medium, the
comoving gauge ($\sv=0$) is identical
to the synchronous gauge ($\alpha=0$), and from the Einstein equation
the comoving-gauge curvature perturbation is conserved $\cp'=0$
\cite{HWNO99R,WASL09}.

Furthermore, in this circumstance, there exists ``Newtonian correspondence'' 
that relates the fully relativistic quantities to the Newtonian quantities.
In a flat universe with
presureless medium, the Newtonian matter density $\dm$ is identical to the
comoving gauge matter density~$\delta_v$, and the Newtonian velocity~$\vN$ and
potential~$\pN$ are identical to the conformal Newtonian gauge quantities
$\vx$ and $\px$ \cite{HWNO99R,HWNO05}. Therefore, we adopt a simple notation
$\dm\equiv\dv$, $\vN\equiv\vx$, and $\pN\equiv\px=-\ax$, but note that
it is fully relativistic and no Newtonian approximation is made.
The Einstein 
and the conservation equations can be written in terms of these quantities:
\beeq
\label{eq:NEW}
\pN={3H_0^2\over2}~{\OM\over ak^2}~\dm ~,\qquad
\vN=-\dm'=-{\HH f\over k^2}~\dm,\qquad
\VGI=-\nhat\cdot\nabla\vN=i\HH f~{\delta_m\over k}~\mu_k~,
\eneq
where the logarithmic growth rate is $f=d\ln\delta_m/d\ln a$ (we used
$d/d\tau=\HH d/d\ln a$), $\mu_k$ is the cosine angle between the 
line-of-sight direction and the wavevector $\mu_k=\kvec\cdot\nhat/k$,
and the equations are in Fourier space.

As the observed galaxy fluctuation in Eq.~(\ref{eq:dobs}) is composed of
many perturbation variables, we first express those perturbation components
in terms of $\dm$, $\vN$, and $\pN$ \cite{YOHAET12,YODE13}
\bear
\label{eq:sum}
\dzg&=&\VGI+\pN+\int_0^{\rbar_z}d\rbar~2\pN'~, \qquad
\drr_\chi=-{\dzg\over\HH}-\int_0^{\rbar_z}d\rbar~2\pN~,\qquad
\kag=-\int_0^{\rbar_z}d\rbar\left({\rbar_z-\rbar\over \rbar\rbar_z}\right)
\hat\nabla^2\pN~,\nonumber \\
&&\hspace{-30pt}
-H{\partial\over\partial\zz}\left({\dzg\over\HH}\right)=-\VGI
-{1+\zz\over H}\pN'-{1+\zz\over H}
{\partial\VGI\over\partial\rbar} -\dzg+{1+\zz\over H}{dH\over dz}~\dzg~,
\quad \dz_{t_p}=\dzg+\HH\vN~, 
\enar
where we have ignored quantities at the observer position
that can be absorbed to the observed mean number density $\widehat{\bar n_g}$
\cite{YOFIZA09,YOO10}. Furthermore, since the observed galaxies are
only those along the past light
cone of photons, the partial derivative with respect to the observed
redshift in Eqs.~(\ref{eq:dobs}) and~(\ref{eq:sum}) is
\beeq
{\partial\over\partial\zz}={1\over H}{d\over d\rbar}=
-{1\over H}\left({\partial\over\partial\tau}
-{\partial\over\partial\rbar}\right)~.
\eneq
The reason for the notation is because we keep the other observable 
quantities $(\ttt,\pp)$ fixed. However, it involves not only the spatial 
derivative, but also the time derivative, as it is literally the variation 
of the observed redshift.

To facilitate the computation, we define the transfer 
functions~$\TT_\PR(k,z)$ that relate the
amplitude of a perturbation variable $\PR(\kvec,z)$ 
of Fourier mode~$\kvec$ at $\zz$
with the initial conditions at very early time. 
In the linear regime, all the perturbations at each wave mode 
grow only in amplitude, without changing its phase set by the initial 
conditions. This deterministic growth of perturbation variables is captured
by the transfer function, and the phase information in the initial
condition is characterized by the curvature perturbation~$\RR(\kvec)$
in the comoving gauge during the inflationary period.\footnote{The 
comoving-gauge curvature perturbation in our notation (Eq.~[\ref{eq:abc}])
is $\cp=\RR$, i.e.,
the curvature perturbation~$\varphi$ in the comoving-gauge condition ($\sv=0$),
but $\RR$ is more commonly used in literature in defining the transfer 
functions. Moreover, it is  sometimes denoted as $\zeta$ in some literature, 
but care must be taken, as $\zeta$ is often used for the curvature 
perturbation $\varphi$ in the uniform-matter gauge ($\delta=0$), i.e., 
$\zeta=\varphi_\delta$ in our notation.}
For example, the 
transfer functions for the matter density fluctuation $\delta_m$ and the
gravitational potential $\pN$ at~$z$ are
\beeq
\label{eq:TT}
\delta_m(\kvec,z)=\TT_m(k,z)\RR(\kvec)~,\qquad
\pN(\kvec,z)=\TT_{\pN}(k,z)\RR(\kvec)=\WM_{\pN}\TT_m(k,z)\RR(\kvec)~,
\eneq
where we defined a conversion function $\WM_\PR$ that relates the transfer 
function~$\TT_\PR$ for a perturbation variable $\PR$
to the transfer function $\TT_m$
for the matter density fluctuation (hence $\WM_m=1$). Other
conversion functions are
\beeq
\label{eq:conver}
\WM_{\pN}={3H_0^2\over2}{\OM\over ak^2}~,\quad
\WM_{\vN}=-{\HH f\over k^2}~,\quad
\WM_\VGI={\HH f\over k^2}{\partial\over\partial\rbar}~,\quad
\WM_{\pN'}=\HH(f-1)\WM_{\pN}~,\quad
\WM_\kag=-\left({\rbar_z-\rbar\over\rbar_z\rbar}\right)
\WM_{\pN}\hat\nabla^2~.
\eneq
Transfer functions of other perturbation variables 
such as $\delta z_\chi$, $\delta r_\chi$, and so on can be computed 
in terms of the conversion functions $\WM_\Upsilon$
in Eq.~(\ref{eq:conver}) by
using the relations in Eq.~(\ref{eq:sum}).

The power spectrum of these perturbation variables $\PR_i(\kvec,z)$ can be
computed as
\beeq
\la\PR_i(\kvec,z_1)\PR_j(\kvec',z_2)\ra=(2\pi)^3\delta^D(\kvec+\kvec')
\TT_{\PR_i}(k,z_1)\TT_{\PR_j}(k,z_2)P_\RR(k)~,
\eneq
where the primordial curvature power spectrum is characterized
by the scalar amplitude $A_s$ and the spectral index~$n_s$ at a pivot
scale $k_0$
\beeq
\label{eq:pri}
\Delta^2_\RR(k)={k^3\over2\pi^2}P_\RR(k)
=A_s\left({k\over k_0}\right)^{n_s-1+{1\over2}{dn_s\over d\ln k}
\ln\left({k\over k_0}\right)}~.
\eneq
Since the observed galaxy fluctuation
in Eq.~(\ref{eq:dobs}) is a linear combination of many perturbation
variables (some of which are also linear combinations or
involve the line-of-sight integration of other perturbation
variables), it proves convenient to write the observed galaxy fluctuation
as
\beeq
\label{eq:decomp3D}
\delta_g^\up{obs}(\bdv{x}_s)=\sum_{\PR_i}\int{d^3\kvec\over(2\pi)^3}
\int_0^{\rbar_s}d\rbar~\WW_i(\rbar)~\TT_{\PR_i}(k,\rbar)\RR(\kvec)~
e^{i\kvec\cdot\bdv{x}}~,\qquad \bdv{x}_s=\rbar_s\nhat~,
\eneq
where $\rbar_s$ is the comoving distance to the source galaxy (i.e., 
$\rbar_s=\rbar_z$), $\bdv{x}=\rbar\nhat$,
and the index-$i$ runs for all the components in Eq.~(\ref{eq:dobs}). 
$\WW_i$ is the Dirac delta function if the perturbation variable $\PR_i$ is
a local function such as $\pN$, $\vN$, and so on, while $\WW_i$ is unity
if it involves the line-of-sight integration such as $\kag$.

\subsection{Galaxy Angular Power Spectrum $C_l$}
\label{ssec:angular}
While galaxy redshift surveys have information on the radial position of
galaxies (based on the observed redshift), two-dimensional angular statistics
may be used to probe cosmology, for example, 
when the radial information is less reliable
due to photometric redshift measurements. By counting the number of galaxies 
$dN_g^\up{obs}(\nhat)$
within the observed solid angle $d\Omega$, the observed angular galaxy
number density $n_{g,\up{2D}}^\up{obs}(\nhat)$ and its fluctuation 
$\delta_{g,\up{2D}}^\up{obs}(\nhat)$ are defined as
\beeq
\label{eq:twod}
dN_g^\up{obs}(\nhat)=\nobsTWO(\nhat)d\Omega~,\qquad
\dobsTWO(\nhat)={\nobsTWO(\nhat)\over\widehat{\bar n_{g,\up{2D}}}}-1~,\qquad
\widehat{\bar n_{g,\up{2D}}}={N_g^\up{tot}\over\Omega}~,
\eneq
where $N_g^\up{tot}$ is the total number of observed galaxies within the
survey area $\Omega$ in angle. The angular fluctuation of the observed
galaxy number density is then decomposed in terms of spherical harmonics as
\beeq
\label{eq:alm}
\dobsTWO(\nhat)=\sum_{lm}a_{lm}Y_{lm}(\nhat)~,\qquad
a_{lm}=\int d^2\nhat~ Y_{lm}^*(\nhat)\dobsTWO(\nhat)~,
\qquad
C_l=\sum_m{\langle a^*_{lm}a_{lm}\rangle\over2l+1}~,
\eneq
and the angular power spectrum $C_l$ 
is the ensemble average of the decomposed angular
coefficients. The angular fluctuation
and the angular power spectrum are constructed purely based on the observable
quantities: the observed angle~$\nhat$ and 
the number of galaxies $dN_g^\up{obs}(\nhat)$.

In order to compare to the observation, we need to compute the 
theoretical predictions
for $\dobsTWO(\nhat)$ and $C_l$. Since the observed
angular galaxy number density $\nobsTWO(\nhat)$ is just the
volume average of the three-dimensional galaxy number density 
$n_g^\up{3D}(\zz,\nhat)$ in Eq.~(\ref{eq:obsng}):
\beeq
\label{eq:2d3d}
\nobsTWO(\nhat)=\int dz~{\rbar^2(z)\over(1+z)^3 H(z)}~n_g^\up{3D}(\zz,\nhat)~,
\eneq
the angular fluctuation of the observed galaxy number density can be related
to the three-dimensional galaxy fluctuation in Eq.~(\ref{eq:dobs}) as
\beeq
\label{eq:PZ}
\dobsTWO(\nhat)=\int dz~\PZ(z)~\dobsTHR(z,\nhat)~,
\qquad 
\PZ(z)={\Omega\over N_g^\up{tot}}{\rbar^2(z)\over(1+z)^3 H(z)} ~
\widehat{\bar n_g}(z)~,
\eneq
where we defined the normalized redshift distribution $\PZ(z)$ of 
the galaxy sample.\footnote{Here we use $n_g^\up{3D}$ instead of 
$n_g^\up{obs}$ (and similarly so for its fluctuation), 
because the three-dimensional quantities may not be available in photometric
surveys to be ``observed.'' In this case, they have to be modeled.} 
It is also noted that the dimensions are different for
the angular and the three-dimensional galaxy number densities.
Therefore, using the decomposition of $\delta_g^\up{obs}(\xvec_s)$ in 
Eq.~(\ref{eq:decomp3D}), 
the observed galaxy angular power spectrum can be computed as
\beeq
\label{eq:cl}
C_l=\langle a^*_{lm}a_{lm}\rangle
=4\pi\int {dk\over k}~\Delta^2_\RR(k)~\TC_l^2(k)~,
\eneq
where we defined the angular multipole function 
\beeq
\TC_l(k)=\sum_{\PR_i}\int dz~\PZ(z)\int_0^{\rbar_z}d\rbar~\WW_i(\rbar)
\TT_{\PR_i}(k,\rbar)~j_l(k\rbar)~.
\eneq
and used the partial wave expansion
\beeq
e^{i\bdv{k}\cdot\bdv{x}}=4\pi\sum_{lm}i^l~j_l(k\rbar)~Y_{lm}^*(\Kang)~
Y_{lm}(\nhat)~,\qquad \hat\nabla^2Y_{lm}(\nhat)=-l(l+1)Y_{lm}(\nhat)~.
\eneq

Accounting for the relativistic effect in galaxy clustering,
the angular power spectrum was computed (see 
\cite{YOFIZA09,CHLE11,BODU11,BRCRET12,MAZHET13}
for plots and detailed explanation). Since the
angular power spectrum is a projected quantity, the line-of-sight velocity
contribution is suppressed in $\dobs$, while the gravitational lensing 
contribution accumulates if the source distribution is located at higher 
redshift. The gravitational potential contribution is small, but is dominant
over other contributions on large scales. Therefore, the relativistic effect
is important on scales $l\sim k/\rbar\sim\HH/\rbar$, but 
 the cosmic variance accordingly grows larger.
The calculation can be readily extended to the angular cross-power spectrum,
in which two different galaxy samples or the same galaxy sample but at
two different redshift distributions are correlated. In particular, the
relativistic effect in galaxy clustering may be isolated with clever choice
of redshift bins  (see \cite{BODU11,DIMOET13,DIMOET14}),
while the detection significance needs to be quantified
with realistic covariance matrix, as the radial bins are not independent.

\subsection{Galaxy Power Spectrum $P_g(k)$}
The initial perturbations generated during the inflation epoch are best
characterized by its power spectrum such as the comoving-gauge curvature
power spectrum $P_\RR(k)$. Its understanding from the galaxy power spectrum
$P_g^\up{obs}(k)$ measurements is one of the main goals in galaxy redshift
surveys, in which galaxy positions are mapped with three-dimensional 
information.  There exist a few complications in predicting the observed
galaxy power spectrum.
Since the power spectrum is inherently non-local, we have to explicitly account
for the survey geometry in consideration. Furthermore, the observed galaxy
fluctuation $\dobs$ in Eq.~(\ref{eq:dobs}) receives contributions from
the fluctuations along the line-of-sight direction, which are nearly 
angular quantities and ill-described by the three-dimensional power
spectrum. For simplicity, we ignore these complications and proceed to
provide a simplified version of the observed galaxy power spectrum
that are just function of local terms in Eq.~(\ref{eq:dobs}).

With this simplification and using the Einstein equation, the observed galaxy
fluctuation in Eq.~(\ref{eq:dobs}) can be expressed in terms of its Fourier
components as
\beeq
\label{eq:Fdec}
\dobs(\zz,\nhat)\approx
\int{d^3\kvec\over(2\pi)^3}~e^{i\kvec\cdot\xvec_s}\left[\delta_g^\up{Newt.}
(\kvec,\zz)+{\cA\over(k/\HH)^2}\dm(\kvec,\zz)
-i\mu_k{\cB\over k/\HH}\dm(\kvec,\zz)\right]~,
\eneq
where the two dimensionless coefficients $\cA$ and~$\cB$ are defined 
as\footnote{It is noted that the coefficient $\cB$ should be distinguished 
from the comoving curvature $\RR(\kvec)$.}
\bear
\label{eq:cab}
\cA&=& e f-{3\over2}\OM(z)\bigg[e+f-{1+\zz\over H}{dH\over dz}
+(t-2)\left(2-{1\over\HH\rbar_z}\right)\bigg] ~,\\
\cB&=&f\left[e-{1+\zz\over H}{dH\over dz}+(t-2)\left(1-{1\over\HH\rbar_z}
\right)\right] ~,\nonumber
\enar
and the Newtonian description of the observed galaxy fluctuation in the
redshift space is \cite{KAISE84}
\beeq
\label{eq:kai}
\delta_g^\up{Newt.}(\kvec,\zz)=b~\dm(\kvec,\zz)-\mu_k^2~
{k^2\vN(\kvec,\zz)\over\HH}=(b+f\mu_k^2)~\dm(\kvec,\zz)~.
\eneq
Apparent from the spatial dependence of $\cB$ and $\cA$, these two new
coefficients in the relativistic description result from the velocity
and the gravitational potential contribution to galaxy clustering
(see \cite{YOHAET12,JESCHI12} for the time-evolution of $\cA$ and $\cB$).
It is noted that these coefficients are derived by assuming general relativity,
and their values and time evolution differ in other gravity theories
\cite{LOYOKO13}. 
The Newtonian description $\delta_g^\up{Newt.}$ is derived from 
Eq.~(\ref{eq:dobs}) by ignoring the gravitational potential 
($\propto\cA\delta_m/k^2$) and the velocity ($\propto\cB\delta_m/k$)
contributions, and the redshift-space distortion term in $\delta_g^\up{Newt.}$
comes from
$-H_z\partial_z(\dzg/\HH)$ in Eq.~(\ref{eq:dobs}).

The observed galaxy power spectrum can be obtained by taking
the ensemble average of the square bracket in Eq.~(\ref{eq:Fdec})
\beeq
P_g^\up{obs}(\kvec,\zz)=P_g^\up{Newt.}(\kvec,\zz)+\Bigg[
{\cA^2\over (k/\HH)^2}+\mu_k^2\cB^2+2b\cA+2f\mu_k^2\cA\Bigg]
{P_m(k,\zz)\over(k/\HH)^2}~,
\eneq
where the Newtonian galaxy power spectrum is computed under the
distant-observer approximation as
\beeq
P_g^\up{Newt.}(\kvec,\zz)=(b+f\mu_k^2)^2P_m(k,\zz)~,\qquad 
P_m(k,\zz)=\TT^2_m(k,\zz)P_\RR(k)~.
\eneq
The relativistic galaxy power spectrum was computed (see
\cite{YOO10,BASEET11,JESCHI12,YOHAET12,LOMERI12,LOYOKO13} for plots and
detailed explanation). While the Newtonian
contribution in Eq.~(\ref{eq:kai}) falls as $k^{n_s}$ on large scales,
the gravitational potential $\cA$ and the velocity $\cB$ contributions 
becomes larger on large scales, as their power spectra scale with
$k^{n_s-4}$ and $k^{n_s-2}$, respectively.
Naturally, these relativistic effects are dominant on large scales, and
hence it is difficult to measure in low-redshift galaxy surveys.
However, with multi-tracer technique \cite{SELJA09}, the cosmic variance,
the dominant source of measurement uncertainties on large scales,
can be removed, and these relativistic effect in galaxy clustering can be
measured with high significance \cite{YOHAET12}. Last, the effect of
the primordial non-Gaussianity is also a relativistic effect in galaxy
clustering and can be readily implemented in the relativistic formula
\cite{YOHAET12,JESCHI12,BRCRET12}.

The power spectrum analysis is often performed by embedding
the observed sphere in a cubic volume and by taking Fourier transformation
of $\dobs$. This procedure practically assumes the distant-observer 
approximation, and the equations in this subsection are 
valid only in the flat-sky limit,
which breaks down on large scales \cite{YOSE13a}.
A more careful power spectrum analysis in observation
can be performed without the distant-observer approximation, but its 
application has been so far limited to the Newtonian redshift-space power 
spectrum \cite{YANAET06}. Despite these shortcomings, the power spectrum 
analysis in this section provides the main relativistic effect in 
galaxy clustering and its detectability in future surveys \cite{YOHAET12}.

\subsection{Spherical Galaxy Power Spectrum $S_l(k)$}
To overcome the shortcoming of the power spectrum analysis, an alternative 
analysis was developed in the past
based on the radial and angular eigenfunctions of the Helmholtz equation
\cite{BIQU91,FILAET95,HETA95},
while its application was limited to the Newtonian expression in 
Eq.~(\ref{eq:kai}). The angular fluctuation in the observed galaxy
fluctuation $\dobs$ is decomposed in terms of spherical harmonics for
all-sky analysis, 
which naturally implements the angular contributions to $\dobs$ such as $\kag$.
The radial fluctuation in $\dobs$ is decomposed in terms of spherical Bessel
function for spectral Fourier analysis.

Using the spherical Fourier analysis, the observed galaxy number density
$n_g^\up{obs}$ can be written in terms of its spherical Fourier mode 
$n_{lm}(k)$ as
\beeq
\label{eq:nlm}
n_g^\up{obs}(\zz,\nhat)
=\int_0^\infty dk\sum_{lm}\sqrt{2\over\pi}~k~j_l(k\rbar_s)~Y_{lm}(\nhat)~
n_{lm}(k)~, \quad
n_{lm}(k)=\int d^3\xvec_s~\sqrt{2\over\pi}~k~j_l(k\rbar_s)~
Y^*_{lm}(\nhat)~n_g^\up{obs}(\xvec_s)~.
\eneq
Since the mean galaxy number density evolves in time, we define the survey
window function $\RW(\zz)$ (or the radial selection function) 
to separate the radial fluctuation from the mean variation as
\beeq
\bar n_g(\zz)\equiv \bar n_g\RW(\zz)~,\qquad 
V_s=4\pi\int dz~{\rbar^2\over H}~\RW(\zz)~,\qquad
\PZ(\zz)={4\pi\over V_s}{\rbar^2\over H}~\RW(\zz)~,
\eneq
where the overall mean number density in the survey is 
$\bar n_g\equiv N_\up{tot}/V_s$. As this radial distribution of the mean number
density only affects the spherical monopole $n_{00}(k)$, the spherical
Fourier mode of the observed galaxy fluctuation and its spherical power
spectrum can be obtained as
\beeq
\delta_{lm}(k)={n_{lm}(k)\over \bar n_g}~,\qquad
\AVE{\delta_{lm}(k)\delta_{l'm'}^*(k')}=\delta_{ll'}\delta_{mm'}\SP_l(k,k') ~.
\eneq
If $n_g^\up{obs}(\zz,\nhat)\propto\dm(\xvec)$, for example, the spherical
power spectrum is then $\SP_l(k)=P_m(k)$, where
$\SP_l(k,k')=\delta^D(k-k')\SP_l(k)$.

Generalizing the spherical Fourier analysis to the relativistic description
in Eq.~(\ref{eq:dobs}), the spherical Fourier mode of~$\dobs$  is
\beeq
\delta_{lm}(k) =i^l\int\!\!{d\ln k'k'^3\over2\pi^2}\int\!\! d^2\Kang'~
\RR(\kvec')~Y_{lm}^*(\Kang')~\ST_l(k',k) ~,
\eneq
and the spherical galaxy power spectrum becomes
\beeq
\label{eq:sss}
\SP_l(k,k')= 4\pi
\int\!\! d\ln\tilde k~\Delta^2_{\RR}\!(\tilde k) \ST_l(\tilde k,k)
\ST_l(\tilde k,k')~,
\eneq
where the spherical multipole function $\ST_l(k',k)$ is defined as
\beeq
\ST_l(k',k)=\sqrt{2\over\pi}
\int_0^\infty d\rbar_s~\rbar_s^2~\RW(\rbar_s)~k~j_l(k\rbar_s)\sum_{\PR_i}
\int_0^{\rbar_s}d\rbar~\WW_i(\rbar,k',l)~\TT_{\PR_i}(k',\rbar)~j_l(k'\rbar)~.
\eneq
The relativistic spherical galaxy power spectrum was computed 
(see \cite{YODE13} for details).
All the relativistic effects in galaxy clustering are naturally implemented
in the spherical Fourier analysis, and the spherical power spectrum
$\SP_l(k)$ reduces to the flat-sky power spectrum $P(k)$ on small scales.
The great advantage in this approach is the spectral Fourier analysis on a
sphere, providing the most natural way to describe
the relativistic effect
in galaxy clustering on large scales. However, a complication is that
the observed data $\dobs(\zz,\nhat)$ needs to be processed as a function
of cosmological parameters, because it requires the conversion of a distance
$\rbar$ to a Fourier mode~$k$, using the observed redshift and 
angle.\footnote{This requirement applies both for $P(k)$ and $\SP_l(k)$ 
analysis, and it can be computationally challenging.} 
The current power spectrum
analysis produces the observed galaxy power spectrum based on the fiducial
cosmology and neglects its model dependence. However, given the measurement
uncertainties in the current surveys, these systematic errors are negligible,
as long as the fiducial model is close to the best-fit cosmology from the
measurements.

\section{Conclusion and Future Prospects}
\label{sec:future}
We have provided a pedagogical derivation of the general relativistic 
description of galaxy clustering and computed the galaxy two-point statistics.
The gauge-invariance of individual equations is explicitly verified to show
that the final relativistic formula for galaxy clustering is indeed 
gauge-invariant. Accounting for the relativistic effect in galaxy clustering,
various galaxy two-point statistics are derived with particular attention
to the relation between various two-point
statistics and the observable quantities.

While the relativistic formula provides the most accurate and complete
description of galaxy clustering on large scales, the linear-order calculation
in this work is limited to the two-point statistics. However, crucial
information about the early Universe is encoded in the deviation from
the Gaussianity, i.e., higher-order statistics such as the bispectrum,
because any deviation from the standard single field inflationary model or
any physics beyond the standard model naturally involves multiple fields,
playing significant roles
at the early Universe. Given the numerous upcoming surveys,
the second-order relativistic description of 
galaxy clustering \cite{YOZA14,YOO14b} 
(see also \cite{BEMACL14,BEMACL14b,DIDUET14})
is an essential tool for probing the subtle relativistic
effect in galaxy clustering that may decode the dynamics of the early Universe.

\acknowledgments
J.~Y. acknowledges useful discussions with Matias Zaldarriaga.
J.~Y. is supported by the Swiss National Science Foundation and the 
Tomalla foundation grants.

\bibliography{ms.bbl}

\clearpage
\begin{table*}
\caption{Various symbols used in the paper}
\begin{tabular}{ccc}
\hline\hline
Symbols & Definition of the symbols & Equation\\
\hline
$g_{ab}$, $\eta_{\mu\nu}$ & FLRW metric \& Minkowsky metric & 
(\ref{eq:bgmetric}) \\
$a$, $\gbar_{\alpha\beta}$
& comoving scale factor \& background three-metric & (\ref{eq:bgmetric})\\
$\AA$, $\BB_\alpha$, $\CC_{\alpha\beta}$
& components of perturbed metric tensor & (\ref{eq:abc})\\
$\alpha$, $\beta$, $\varphi$, $\gamma$, $B_\alpha$, $C_\alpha$ 
& decomposed metric perturbations & (\ref{eq:abc})\\
$\UU^a$, $\VV^\alpha$, $\sv$ & four velocity, spatial component of $\UU^a$, 
scalar velocity & (\ref{eq:uuu}) \\
$\xi^a$, $T$, $L$, $L^\alpha$ 
& coordinate transformation vector \& its decomposition & (\ref{eq:gt})\\
$\ax$, $\px$, $\vx$, $\dv$ & scalar gauge-invariant variables &
(\ref{eq:sgi})\\
$\pv_\alpha$, $\gv_\alpha$ & vector gauge-invariant variables 
& (\ref{eq:vgiv})\\
$k_L^a$, $k^a$ & photon wavevector in local \& FRW frames
& (\ref{eq:Lwave}), (\ref{eq:Fwave}) \\
$\ttt$, $\pp$, $\zz$
& observed angular position $\nhat$
\& redshift of source galaxy & (\ref{eq:oang}),
(\ref{eq:zrat})\\
$[e_t]^a$, $[e_i]^a$ & tetrad vectors in local frame
& (\ref{eq:tetrad})\\
$\oo$, $\cc$ & physical \& conformal affine parameters & (\ref{eq:af})\\
$\CK^a$ & conformally transformed photon wavevector & (\ref{eq:Cwave})\\
$\dnu$, $\dea^\alpha$ & perturbations in photon wavevector $\CK^a$
& (\ref{eq:Cwave})\\
$\dnug$, $\deag$ & gauge-invariant variables for $\dnu$, $\dea^\alpha$ 
& (\ref{eq:dnugi})\\
$\hat x^a_s$, $x^a_s$ &
observationally inferred \& true positions of source galaxies 
& (\ref{eq:inferred}), (\ref{eq:disp}) \\
$\bar\tau$, $\bar r$ & comoving coordinates in the background
& (\ref{eq:bar}) \\
$\drr$, $\dtt$, $\dpp$  & distortions in the source position
between $\hat x^a_s$ \& $x^a_s$ & (\ref{eq:disp})\\
$\dz$ & distortion in the observed redshift & (\ref{eq:Obsz}),
(\ref{eq:zobs}) \\
$\kappa$ & gravitational lensing convergence & (\ref{eq:kappadef}),
(\ref{eq:kappa}) \\
$\ddL$ & fluctuation in luminosity distance & (\ref{eq:ddl})\\
$\drr_\chi$, $\kag$ & gauge-invariant variables for $\drr$ \& $\kappa$
& (\ref{eq:rg})\\
$dV_\up{phy}$, $d\bar V_\up{obs}$, $\dV$ 
& physical \& observationally inferred volumes, and their difference & 
(\ref{eq:dV}), (\ref{eq:ddV}) \\
$n_g$, $n_g^\up{obs}$ & physical \& observed galaxy number densities 
&(\ref{eq:obsng})\\
$\bar n_g$, $\widehat{\bar n_g}$ & physical \& observed mean number densities
& (\ref{eq:png}), (\ref{eq:mmm})\\
$\delta_g^\up{obs}$, $\delta_g^\up{int}$ & observed \& intrinsic galaxy
fluctuations & (\ref{eq:dobsdef}), (\ref{eq:png}) \\
$\pN$, $\vN$, $\delta_m$ & Newtonian counterparts of $\px$, $\vx$, $\dv$
& (\ref{eq:NEW}) \\
$\TT_\Upsilon$, $\WM_\Upsilon$ & transfer function \& its conversion
function to $\TT_m$ & (\ref{eq:TT}), (\ref{eq:conver}) \\
$\Delta^2_\RR$ & dimensionless power spectrum of primordial curvature
perturbation & (\ref{eq:pri})\\
$\nobsTWO$, $\dobsTWO$ & two dimensional galaxy
number density \& its fluctuation & (\ref{eq:twod})\\
$\PZ$ & normalized redshift distribution of source galaxies & (\ref{eq:PZ})\\
$C_l$, $\TC_l$ & angular power spectrum \& its multipole function
& (\ref{eq:cl}) \\
$\cA$, $\cB$ & gravitational \& velocity contributions to $\dobs$
& (\ref{eq:cab}) \\
$n_{lm}(k)$, $a_{lm}$ & spherical \& angular decompositions & (\ref{eq:nlm}),
(\ref{eq:alm}) \\
$\SP_l$, $\ST_l$ & spherical power spectrum \& its multipole function
& (\ref{eq:sss}) \\
\hline\hline
\end{tabular}
\label{tab:tab}
\end{table*}

\end{document}